\documentclass{PoS}
\usepackage{amsmath,slashed,subfigure,setspace}
\usepackage{epstopdf}
\pdfoutput=1
\usepackage[utf8]{inputenc}
\usepackage[T1]{fontenc}
\newcommand{\su}[1]{\ensuremath{\text{SU}#1}}

\newcommand{\gino}{\lambda}

\newcommand{\Nsusy}{\mathcal{N}}

\newcommand{\ps}{\psi}
\newcommand{\psb}{\bar{\psi}}
\newcommand{\arxiv}[1]{arXiv:\,\href{http://arxiv.org/abs/#1}{{\tt #1}}}

\title{Supersymmetric and conformal theories on the lattice: from super Yang-Mills towards super QCD}

\ShortTitle{Supersymmetric and conformal theories}

\author{\speaker{Georg Bergner}\\
Friedrich-Schiller-University Jena, Institute of Theoretical Physics,\\
Max-Wien-Platz 1, D-07743 Jena, Germany\\
E-mail: \email{georg.bergner@uni-jena.de}}

\author{Stefano Piemonte\\
Universit\"at Regensburg, Institute for Theoretical Physics,\\
D-93040 Regensburg, Germany\\
E-mail: \email{stefano.piemonte@ur.de}}

\abstract{%
This talk is an overview of our recent investigations of supersymmetric and near conformal gauge theories. We have studied extensively $\Nsusy=1$ super Yang-Mills theory, most recently with the gauge group \su{(3)}. In addition we have investigated theories that show indications for a conformal behavior at an infrared fixed point. We have included a mixed fundamental and adjoint fermion action setup in our studies. I will explain how this is related to the investigation of supersymmetric QCD on the lattice and present some first studies of the main obstacles that need to be addressed in the investigation of this theory. }

\FullConference{The 36th Annual International Symposium on Lattice Field Theory - LATTICE2018\\
		22-28 July, 2018\\
		Michigan State University, East Lansing, Michigan, USA.}

\begin{document}

\section{Investigations of supersymmetric Yang-Mills theory and near conformal theories}
In recent years our collaboration has published a number of papers on two interesting related subjects. The first one is the study of supersymmetric Yang-Mills theory, the pure gauge sector of 
supersymmetric extensions of the Standard Model \cite{Bergner:2015adz,Ali:2018dnd,Ali:2018fbq}. The second one is the investigation of the strong dynamics of composite Higgs and walking Technicolor theories \cite{Bergner:2017gzw,Bergner:2017ytp,Bergner:2016hip}. Both of these projects are connected to 
the new studies of supersymmetric QCD that are reported in this contribution.

There are basically two alternative motivations for the interest to lattice simulations of supersymmetric strongly interacting theories. On the one hand there are the investigations in the context of possible 
extensions of the Standard Model based on a supersymmetric completion. The supersymmetric extension of strong interactions corresponds to supersymmetric QCD. In the gauge sector, fermion partners, the gluinos, are added to the boson gluon fields; in the matter sector, the squarks are added as boson counterparts of the quark fields. A strongly interacting sector might also explain the supersymmetry breaking terms that are required 
to make the supersymmetric extensions of the Standard Model consistent with the experimental observations.

On the other hand there is a general theoretical interest in supersymmetry that is a tool providing a better understanding of the non-perturbative dynamics of certain theories based on dualities. A well-known example is $\mathcal{N}=4$ supersymmetric Yang-Mills theory, but also $\mathcal{N}=2$ supersymmetric Yang-Mills and $\mathcal{N}=1$ supersymmetric QCD are interesting from this perspective. It is our project to test these phenomena with numerical lattice simulations.

In a quite similar way, there are two basic motivations for the studies of conformal and near conformal strongly interacting theories. The first one is the phenomenological possibility of a new strong dynamics beyond the Standard Model that plays the role of the Higgs sector.
It turned out that theories with a near conformal, ``walking'', behavior can accommodate large scale separation between the scale of the electroweak symmetry breaking and the scale of a strongly interacting completion inducing the effective operators for fermion mass generation. In this way the tension between the experimental constraints and the extensions of the Standard Model can be released. Interesting candidates for such a walking dynamics are theories with fermions in the adjoint representation which are closely related to supersymmetric theories. 
The most prominent example of a candidate for a walking scenario is the \su{(2)} gauge theory with two Dirac fermions (equivalent to four Majorana) in the adjoint representation, called Minimal Walking Technicolor, whose conformal behavior has been investigated with several different numerical methods on the lattice. A study of a theory with three adjoint Majorana flavours, which completes our scan in the fermion number for adjoint QCD, is presented in a separate contribution to this conference \cite{nf32Lat18}.

The second motivation for the studies of (near) conformal strongly interacting theories are again more general theoretical questions. The perturbative $\beta$-function develops a fixed point when the number of fermion flavors is increased. A second zero of the $\beta$-function means that there is a window for theories with a non-perturbative conformal behavior at an infrared fixed point below the loss of asymptotic freedom, at an even larger number of fermions. The perturbative analysis needs to be verified by non-perturbative methods. If such a behavior is confirmed, it raises additional questions concerning the non-QCD-like dynamics of conformal and near conformal theories. Of particular interest are the effective field theory descriptions in the near conformal case, where a light scalar bound state, a dilaton, might be introduced in contrast to standard chiral perturbation theory for QCD. Since the non-perturbative dynamics of the near conformal theories is only accessible by numerical methods and subject to unavoidable uncertainties, the near conformal or conformal nature of a number of theories is still under debate. It might therefore provide a more solid ground for the theoretical analysis if the near conformal nature of a theory would be known by some alternative analytical arguments.

\section{Supersymmetric Yang-Mills theory and supersymmetric QCD}
$\mathcal{N}=1$ supersymmetric Yang-Mills theory (SYM) is the pure gauge part of supersymmetric QCD, with the gluino fields $\lambda$ and the usual gauge interactions,
\begin{equation}
 \mathcal{L}_{SYM}=\frac{1}{4}
F_{\mu\nu}F^{\mu\nu}+\frac{1}{2}\bar{\gino}(\slashed{D}+m_g)\gino\; .
\end{equation}
A non vanishing gluino mass $m_g$ leads to a soft supersymmetry breaking. The gluinos are Majorana fermions in the adjoint representation.

In separate contributions to this conference the status of our investigations of SYM for the gauge group \su{(2)} and \su{(3)} is reported~\cite{BaryonLat18}. The basic investigations of 
these theories are now nearly completed. In a second contribution the status of our analysis of phase transitions in SYM is presented, in particular regarding the measurement of the gluino condensate with the methods of the Gradient flow~\cite{GflowLat18}.

The next natural step is the introduction of an $N_c\oplus\bar{N}_c$ chiral matter superfield consisting of a Dirac quark $\psi$ and squarks $\Phi_1$, $\Phi_2$, which are complex scalars.
The Lagrangean is extended by Yukawa and four squark interactions,
\begin{align}
\mathcal{L}_{SQCD}=&\mathcal{L}_{SYM}+|D_\mu \Phi_1|^2+|D_\mu \Phi_2^\dag|^2+\psb(\gamma_\mu D^\mu_f+m)\ps+m^2|\Phi_1|^2+m^2|\Phi_2|^2\\
&+ i\sqrt{2}g \bar{\lambda}^a\left(\Phi_1^\dag   P_{+}+\Phi_2P_- \right)T^a  \psi
 -i\sqrt{2}g \bar{\psi}T^a \left(P_-  \Phi_1 + P_+\Phi_2^\dag \right) \lambda^a\\
&+\frac{g^2}{2} \left( \Phi_1^\dag T^a \Phi_1-\Phi_2 T^a \Phi_2^\dag\right)^2.
\end{align}
Further flavors can be added in terms of additional matter multiplets (quark and squark fields), coupled in the same way to the gluino and gluon fields. The resulting theory is \su{(N_c)} supersymmetric QCD (SQCD) with $N_f$ matter flavors.
This theory has been considered in a perturbative setup on the lattice in \cite{Costa:2017rht} and the general form of possible supersymmetry breaking counter-terms as well as the symmetries have been discussed 
in \cite{Giedt:2009yd}. A perturbative setup for the tuning of the parameters is presented in another contribution to this conference~\cite{WelleLat18}.

There are a number of interesting theoretical predictions for the low energy effective theory and the phase structure of SQCD as a function of $N_c$ and $N_f$ \cite{Intriligator:1995au}.
For $N_c>N_f$ there are no vacuum states in the chiral limit and the vacuum expectation value of the scalar condensate diverges.
In the region $\frac{3}{2}N_c<N_f<3N_c$ the theory is expected to have an infrared fixed conformal point. 
In between these two cases there are theories with chiral symmetry breaking and confinement in the chiral limit.

The prediction of the conformal window and the phase structure of SQCD are quite remarkable findings and it would be interesting to check whether they can be confirmed with numerical non-perturbative simulations.
Lattice simulations of supersymmetric QCD would also open the way for a number of additional investigations, like the dualities of the theory and possible supersymmetry breaking scenarios.

In practice, there are several difficult obstacles that one has to circumvent in order to simulate supersymmetric QCD on the lattice. Supersymmetry is broken in any kind of local lattice realization of the theory.
Therefore the first problem is the tuning towards the supersymmetric continuum limit. In supersymmetric Yang-Mills theory there is only the tuning of a single parameter required, but as soon as the scalar squark fields 
of the matter multiplets are added, a large number of scalar self interactions and Yukawa couplings appear.

The second obstacle for the lattice simulations are the different phases of the theory sketched above. At small $N_f$, the theory has flat directions and a divergent scalar expectation value. Even though there is a clear prediction for this phenomenon, it makes the simulations 
of the theory rather difficult towards the chiral limit. It seems that at small $N_f$ the theory is either close to the heavy mass pure \su{(N_c)} supersymmetric Yang-Mills limit or in a Higgs phase resembling an \su{(N_c-N_f)} supersymmetric Yang-Mills theory except for some small regions of the parameter space.
At $\frac{3}{2}N_c<N_f<3N_c$, the investigations of a conformal theory on the lattice have their own difficulties, but a number of tools have been derived for theories in such a regime and it is interesting to cross-check the 
lattice simulations with the analytic predictions. The theories with $N_c\leq N_f$, but below the conformal window, might in practice be hard to distinguish from the conformal scenario. Supersymmetric QCD is hence always in a regime where the chiral limit is hard to control.

The third obstacle is related to the Yukawa couplings and the chiral symmetry of the theory. In order to have a clear tuning of the scalar and pseudoscalar Yukawa interactions as well as the mass terms, a well defined chiral symmetry on the lattice would be helpful. This can be 
achieved with Ginsparg-Wilson fermions on the lattice. In this case even an offline approach for the tuning is possible as described in \cite{Giedt:2009yd}. However, the computational costs of these fermion realizations is much higher than for the improved Wilson type of fermions
that we have used in our investigations of supersymmetric Yang-Mills theory.

An additional obstacle for the numerical simulations is the sign problem that appears in supersymmetric QCD. The integration of the Majorana fermions yields a Pfaffian. In supersymmetric Yang-Mills theory it is real and positive in the continuum limit, but there can be a contribution
of negative signs on the lattice. In supersymmetric QCD, the Pfaffian is complex and a severe sign problem might appear towards the chiral limit. The imaginary part of the Pfaffian changes sign under a transformation of the scalar fields that leaves the rest of the action invariant.
This means that the Pfaffian occurs with the same probability as its complex conjugate and the average contribution becomes real. The real contribution can in principle be negative if the mass and the scalar field expectation value are small. This situation is most likely not so much relevant at smaller $N_f$, except for 
possible small regions of the parameter space: either squark mass is large or the scalar field gets large expectation values. 

We are addressing these problems first in a simplified setup, investigating the two-flavor \su{(2)} supersymmetric QCD without scalar fields. This should help us to find out whether we are already close to a conformal behavior. 
This theory is interesting in its own right since it is a candidate for a composite Higgs extension of the Standard Model. It is a slightly modified version of the theory proposed in~\cite{Ryttov:2008xe}. 

In the end we discuss also some first results of our simulations of one-flavor supersymmetric QCD. We are currently not able to perform the full non-perturbative tuning of the theory, but we tested an approach that is based on the perturbative matching of scalar and fermion propagators.

\section{Mixed representation setup and the near conformal dynamics of Ultra Minimal Walking Technicolor}
\begin{figure}
\subfigure[comparison to $N_f=2$ fundamental\label{fig:nf2comparison}]{\includegraphics[width=0.45\textwidth]{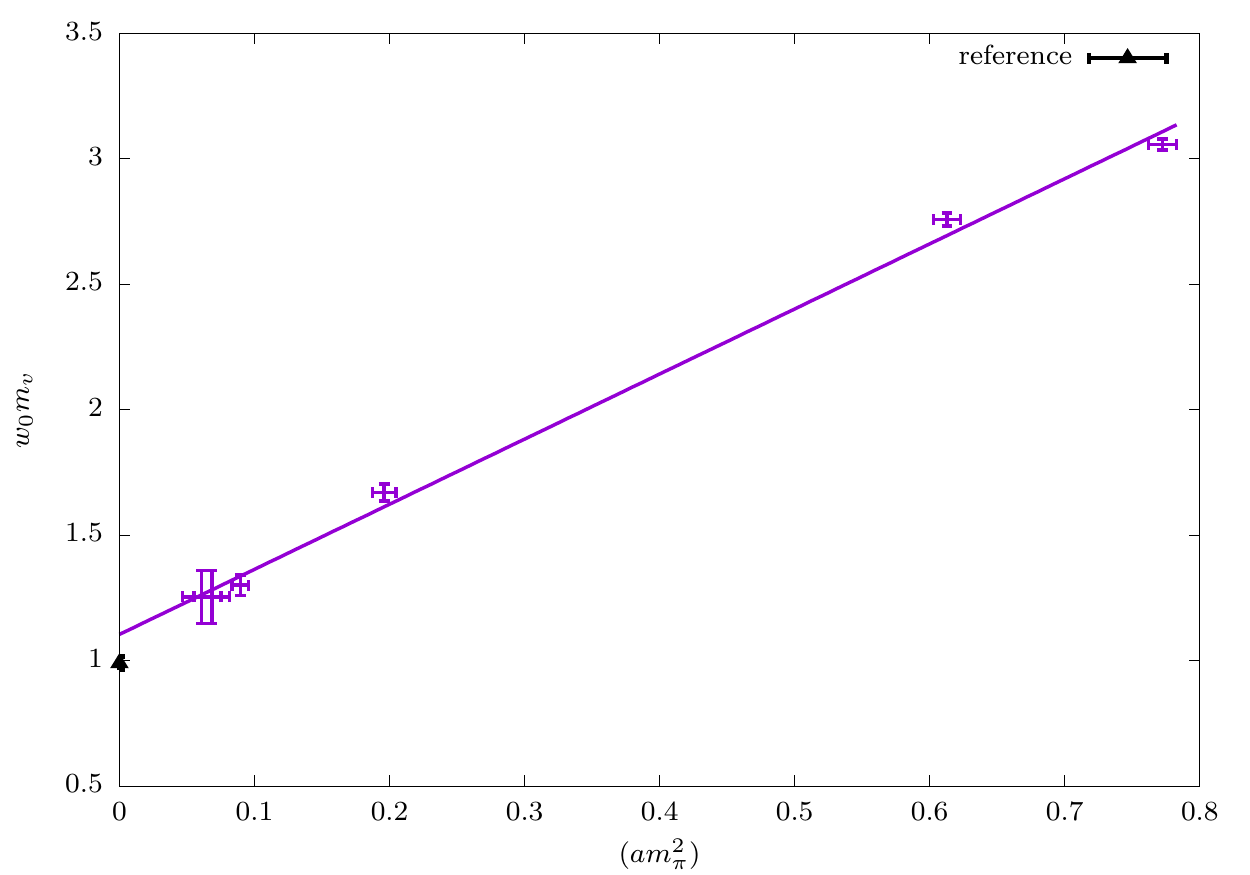}}\quad
\subfigure[fundamental meson mass ratios\label{fig:massratios}]{\includegraphics[width=0.45\textwidth]{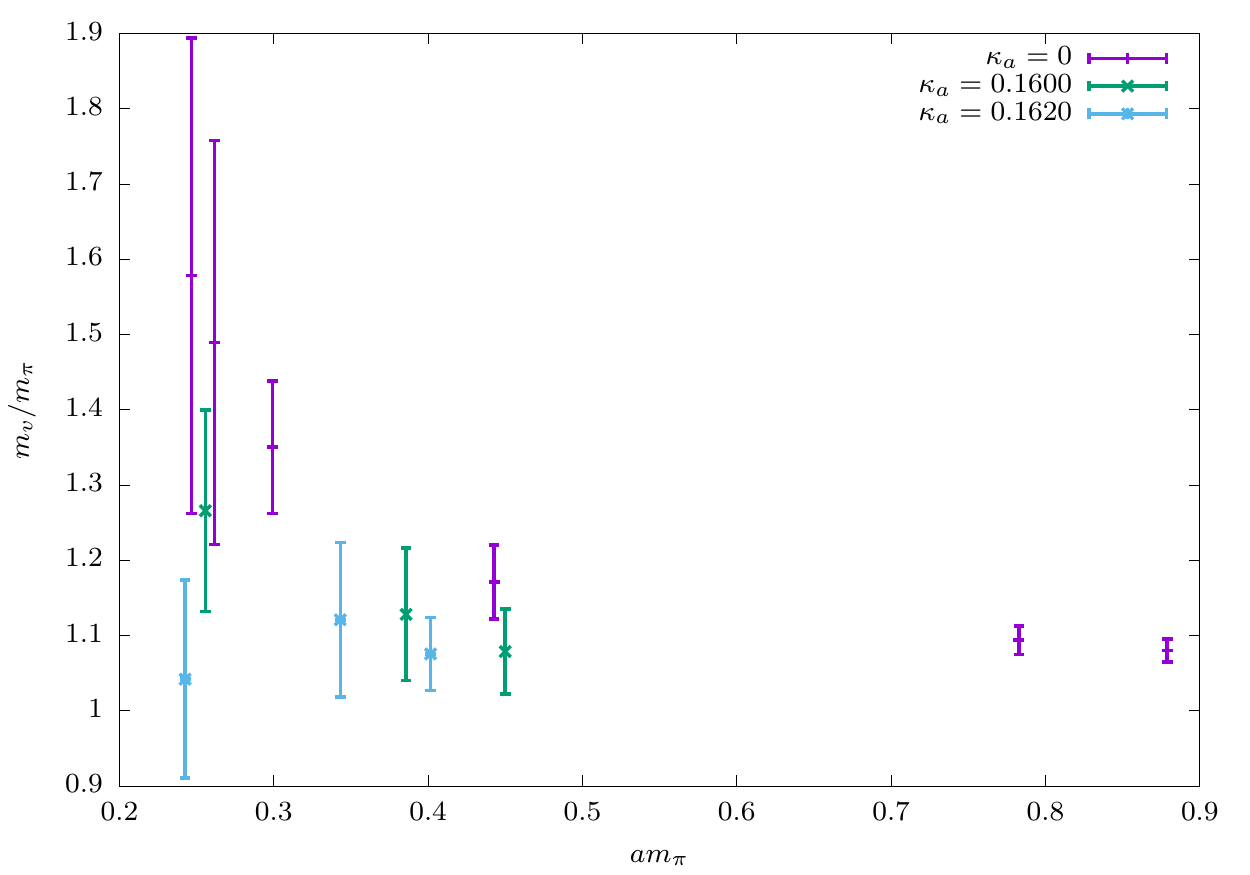}}
\caption{(a) The comparison of the chiral extrapolation of the vector meson mass in units of the Gradient flow scale $w_0$ at $\beta=2.1$ and the results obtained in~\cite{Arthur:2016dir}. Note that the extrapolation of the reference includes a chiral and continuum extrapolation up to quadratic corrections which explains the remaining mismatch. (b) The deviation of the particle spectrum once the Majorana fermion in the adjoint representation is added. The mass ratios of the vector meson and the pion tend towards a constant value.}
\label{fig:umwt1}
\end{figure}
\begin{figure}
\subfigure[adjoint pseudoscalar meson mass as a function of the fundamental mass parameter\label{fig:masstuning1}]{\includegraphics[width=0.45\textwidth]{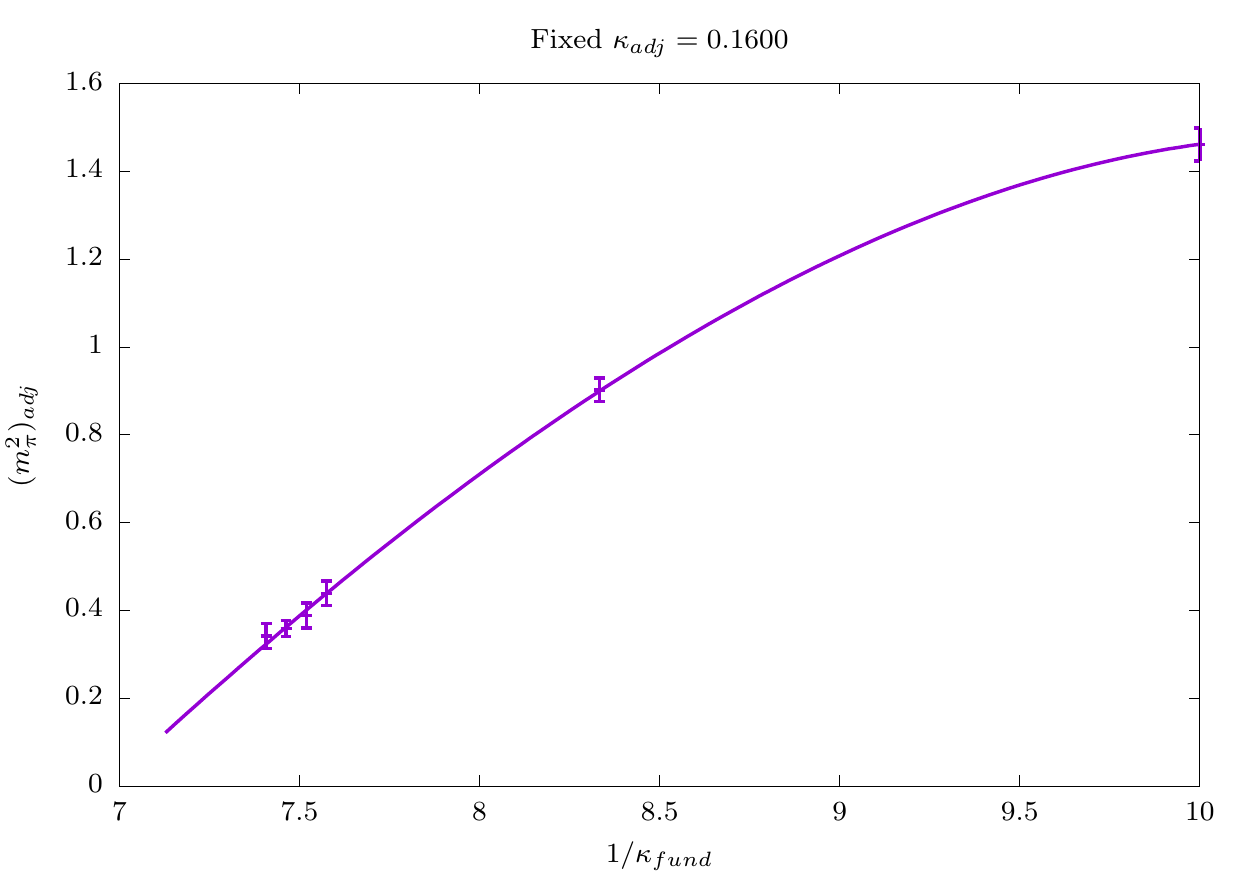}}\quad
\subfigure[fundamental pseudoscalar meson mass as a function of the adjoint mass parameter\label{fig:masstuning1}]{\includegraphics[width=0.45\textwidth]{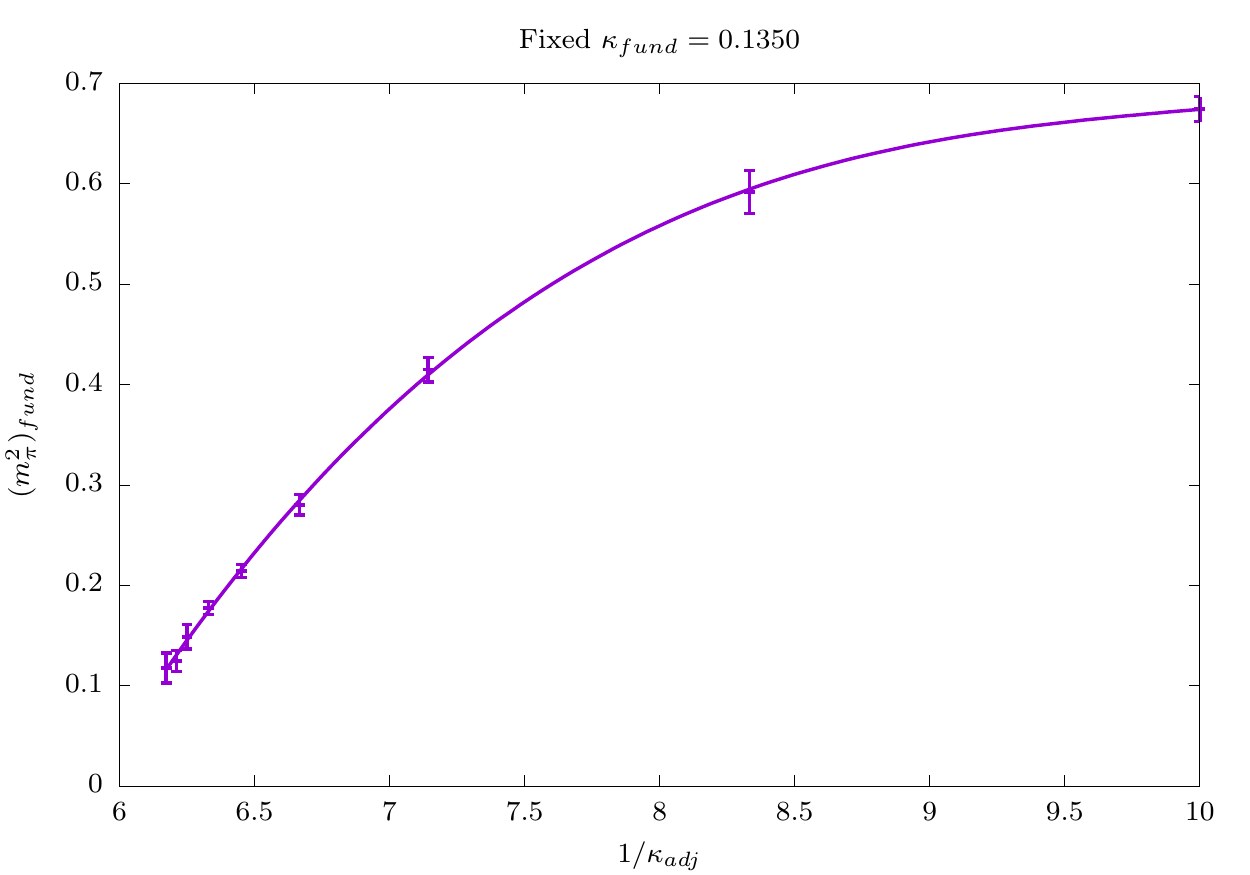}}
\caption{Interplay between the tuning of the fundamental ($\kappa_{fund}$) and adjoint ($\kappa_{adj}$) mass parameters for the simulations at $\beta=2.1$. (a) The dependence of the pion mass for the fundamental representation on the tuning of the adjoint mass parameter. (b) The dependence of the pion mass for the fermions in the adjoint representation on the fermion mass parameter for the fundamental representation.}
\label{fig:umwt2}
\end{figure}
We consider \su{(2)} gauge theory with two Dirac fermions in the fundamental and one Majorana fermion in the adjoint representation. The motivations for the consideration of the theory, besides the mentioned relation to SQCD, are based on the walking Technicolor scenario. In this scenario the Standard Model is extended with a near conformal, or walking, gauge theory in order to accommodate the fermion mass generation with the constraints from electroweak precision measurements. According to naive estimates, Ultra Minimal Walking Technicolor provides a setup with the least tension with the electroweak precision data for a near conformal theory, for a review of other interesting properties of the theory like the possible dark matter candidate see~\cite{Ryttov:2008xe}. This theory has two Dirac fermions in the fundamental and one Dirac fermion in the adjoint representation. Investigations of other theories with fermions in the adjoint representation have shown that the near conformal behavior appears already at a smaller number of fermions than indicated by the naive estimates.

A second important condition for the walking scenario is the large mass anomalous dimension. Minimal Walking Technicolor does not fulfill this condition since a small mass anomalous dimension of $\gamma_\ast=0.2-0.4$ has been found, see \cite{Patella:2012da,Bergner:2016hip} and references therein. It might still be that the complete theory coupled to the Standard Model in an Extended Technicolor setup has slightly different properties, but according to the standard arguments one has to consider alternative candidates. One of the largest mass anomalous dimensions has been found for the \su{(2)} gauge theory with one Dirac fermion in the adjoint representation~\cite{Athenodorou:2014eua}. As expected, the mass anomalous dimension for the adjoint representation increases towards a smaller number of fermions, as confirmed by our comparison of different number of flavors. However, the theory with one Dirac flavor in the adjoint representation has not the right particle content to be combined with the Standard Model. Ultra Minimal Walking Technicolor is the simplest extension of the theory in order to allow the coupling to the Standard Model sector. 

The investigations of Minimal Walking Technicolor and related theories with fermions in the adjoint representation show that the near conformal behavior with a large mass anomalous dimension might appear already for a smaller number of fermion fields than expected from the naive perturbative arguments.
The near conformal behavior, which is hard to distinguish from the conformal one, might appear already for the theory with one Majorana fermion in the adjoint and two Dirac fermions in the fundamental representation. The aim of our investigations is to quantify how close this theory is to a conformal behavior.

These investigations are also an extension of the investigation of \su{(2)} gauge theory with two Dirac fermions in the fundamental representation presented in~\cite{Arthur:2016dir,Drach:2017btk}. The extension with the adjoint fermions is complementary to the studies with additional fundamental fermions~\cite{Karavirta:2011zg}.

The first step in our investigation has been to start from the heavy adjoint flavor limit and cross-check previous results for two fundamental Dirac fermion flavors with our setup. We have used a one-loop clover improved Wilson fermion action and a plain Wilson gauge action. In the limit of a decoupled adjoint flavor, we are able to reproduce the results presented 
in~\cite{Arthur:2016dir,Drach:2017btk} up to a reasonable accuracy as shown in Figure~\ref{fig:nf2comparison}. 

The next step is to resolve the two mass parameters tuning that is required for the simulations in the mixed representation setup. It turns out that there is a strong interplay between of the tuning of the different representations towards the chiral limit. The tuning of the mass for one representation strongly affects the pion mass in the other representation, as shown in Figure~\ref{fig:umwt2}.

As a last step, we have investigated the influence of the fermions in the adjoint representation on meson mass ratios in the fundamental one. The near conformal behavior predicts constant mass ratio of the vector meson mass ($m_v$) over the pseudoscalar (pion) mass ($m_\pi$). In the standard chiral symmetry breaking scenario, which should apply for the case without the additional adjoint Majorana fermion, this ratio diverges in the chiral limit. Our first very preliminary data show the transition from the chiral symmetry breaking scenario towards a conformal one, see Figure~\ref{fig:massratios}.

\section{Numerical simulations of one-flavor supersymmetric QCD}
We have started our first studies of supersymmetric QCD. Besides the additional complication of introducing the additional scalar fields, it requires the implementation of the very special mixed adjoint Majorana / fundamental Dirac Yukawa couplings. We have developed the RHMC algorithm for the simulation of these types of couplings, neglecting at the moment the complex phase of the Pfaffian. In order to be far from a possible conformal behavior, we have considered the one-flavor case for the gauge group \su{(2)}. Our lattice realization employs the same discretization of the derivative operator for the squark and the quark fields, which means a Wilson like derivative term including next-to-nearest neighbor interactions and a Wilson mass for the scalar fields. It implies unbroken supersymmetry in the perturbative limit, setting the gauge coupling to zero. So far we have only considered an unimproved Wilson fermion setup and our results are still very preliminary.
In our first investigations we have set the gauge and Yukawa couplings to their tree-level relations and neglected any further tuning of the supersymmetry breaking operators. One problem that we have to face is the clear signature of the divergent scalar vacuum expectation value. In a scan of the adjoint and fundamental mass parameters, one clearly observes regions with large values for the scalar fields. These regions seem to be close to the expected small fermion mass limit. In the full tuning approach of the parameters this problem has to be controlled in a proper way and we are currently working on a better algorithm for this regime.

\section{Conclusions and outlook}
We have considered in our simulations supersymmetric QCD and related theories in numerical lattice simulations. These theories are interesting from several perspectives: One of them is the richness of non-perturbative phenomena in these theories that can be probed with numerical lattice simulations. The other is a possible deeper understanding of supersymmetric extensions of the Standard Model.

We have started our investigations with a related theory without scalar fields. This theory is an interesting candidate for a composite Higgs (walking Technicolor) extension of the Standard Model. We were able to match previous results for the \su{(2)} gauge theory with two fundamental flavors and investigated the tuning of the mass parameters with two different representations. Our first preliminary results indicate a transition form a chiral symmetry breaking towards a conformal scenario. 

We have also done our first simulations of supersymmetric QCD. Our current setup is motivated by the perturbative limit and a much larger effort is required for the non-perturbative tuning. Besides the tuning to the supersymmetric limit, other problems have to be solved in the simulations of the theory. One peculiar problem is related to the divergent scalar field contributions in the chiral limit. 

Note that the simulations of SQCD must be in a regime where the supersymmetry breaking of the pure Yang-Mills sector can be under control, which means at least lattice sizes of $24^3\times 48$ for unimproved Wilson fermions. Consequently the current simulations for SQCD are still not in the regime where a non-perturbative tuning can be expected to work. A much larger effort is needed to get to the required regime where the supersymmetry breaking can be brought under control.

\begin{spacing}{0.85}
\section*{Acknowledgments}
We thank the members of the DESY-M\"unster collaboration for helpful discussions, comments, and advises.
The authors gratefully acknowledge the Gauss Centre for Supercomputing
e.~V.\,\linebreak(www.gauss-centre.eu) for funding this project by providing
computing time on the GCS Supercomputer JUQUEEN and JURECA at J\"ulich
Supercomputing Centre (JSC) and SuperMUC at Leibniz Supercomputing Centre
(LRZ). G.~B.\ acknowledges support from the Deutsche Forschungsgemeinschaft (DFG) Grant
No.~BE 5942/2-1.

\end{spacing}
\end{document}